\documentclass[12pt]{article} 
\usepackage[dvips]{graphics}
\title{The Lorentz Force Law and Spacetime Symmetries}  
\author{{\it Richard Shurtleff~}\thanks{affiliation and mailing 
address: Department of Applied Mathematics and Sciences, 
Wentworth Institute of Technology, 550 Huntington Avenue, 
Boston, MA, USA, ZIP 02115, telephone number: (617) 989-4338, fax 
number: (617) 989-4591 , web: http://ox.wit.edu//~shurtleffr/ , e-mail address: shurtleffr@wit.edu}} 
\begin{document} 
          
\maketitle 

\begin{abstract} Assume that arc length is measured with the flat spacetime metric. Then, for the most general Poincar\'{e} group representation for translating 4-vectors, curves with parallel translated tangent vectors must have accelerations that are the scalar product of the tangent vector with an antisymmetric tensor. Such curves are the paths of charged particles in an electromagnetic field.

\vspace{0.5cm}

[The New England Sections of the American Physical Society and the American Association of Physics Teachers Spring Meeting took place April 1 and 2 at MIT in Boston. I wish to thank the organizers for allowing me to present this article in the Poster Session at the Stata Center. The abstract and references and Appendices \ref{VTM} through \ref{Ptrans} were posted. Appendix \ref{Verbal} includes explanations presented during the poster session and was written after the conference.]
\end{abstract}

\pagebreak

Successive infinitesimal rotations about the origin, boosts, and translations preserve arc lengths calculated with the flat spacetime metric. The arc length formula is
\begin{equation} \label{ARC2}
     (d\tau)^2 =  -\eta_{\alpha \beta} dx^{\alpha}dx^{\beta} = -(dx^{1})^{2}-(dx^{2})^{2}-(dx^{3})^{2}+(dx^{4})^{2}\quad ,
\end{equation}
where $d \tau$ is the arc length for the interval $dx^{\alpha}$ with $\alpha \in$ $\{1,2,3,4\}$ with indices $\{1,2,3\}$ for rectangular space coordinates and $\alpha$ = 4 indicating a time component. The flat spacetime metric is the diagonal matrix $\eta$ = diag$\{+1,+1,+1,-1\}.$ For time-like intervals, $d\tau$ is the `proper time.'

The transformations make up the group of spacetime symmetries connected to the identity, often called the `Poincar\'{e} group'. Let $\Lambda$ be a transformation that results from successive rotations and boosts and let $\delta x$ be a displacement. Any Poincar\'{e} transformation $(\Lambda,\delta x)$ can be given as a `Lorentz' transformation $\Lambda$ followed by a translation along $\delta x.$ 

A scalar field $f(x^{\alpha})$ changes with a Poincar\'{e} transformation $(\Lambda,\delta x)$ because the event $x^{\alpha}$ acquires the new coordinates $\Lambda^{\alpha}_{\sigma} x^{\sigma} + \delta x^{\alpha},$
$$ U(\Lambda,\delta x)f(x^{\alpha})U^{-1}(\Lambda,\delta x) = f(\Lambda^{\alpha}_{\sigma} x^{\sigma} + \delta x^{\alpha}) \quad . $$
This is the `differential' representation of the Poincar\'{e} group in which the momentum operator is proportional to the gradient.

A field $\psi_{n}(x^\alpha)$ with `spin' has more than one component $n \in$ $\{1,\ldots, N\},$ each component itself a field. A Poincar\'{e} transformation of $\psi_{n}(x^\alpha)$ combines the differential representation with a non-unitary matrix representation, \cite{Tung1, WeinbergFields}
 $$ U(\Lambda,\delta x)\psi_{n}(x^\alpha)U^{-1}(\Lambda,\delta x) = D^{-1}_{n \bar{n}}(\Lambda,\delta x) \psi_{\bar{n}}(\Lambda^{\alpha}_{\sigma} x^{\sigma} + \delta x^{\alpha}) \quad , $$
where $D(\Lambda,\delta x)$ is an $N \times N$ matrix representing the transformation $(\Lambda,\delta x).$ 

In this paper, we consider the most general matrix translations $D(1,\delta x)$ allowed to 4-vectors. The representation involves a direct sum vector-plus-tensor field, an $N$ = 20 component field $\psi_{n}(x)$ with 4 components for the vector field $v^{\mu}(x)$ and 16 components for the second rank tensor field $T^{\mu \nu}(x),$
$$ \psi_{n}(x) = \pmatrix{v^{\mu}(x) \cr T^{\mu \nu}(x)} \quad . $$
The $20 \times 20$ matrix representation $D(\Lambda,\delta x)$ is described in Appendix \ref{Poincare}. 

\pagebreak

Applying a translation to $\psi_{n}(x)$ yields, by Appendix \ref{Poincare},
$$ U(1,\delta x)\psi_{n}(x^\alpha)U^{-1}(1,\delta x) = D^{-1}_{n \bar{n}}(1,\delta x) \psi_{\bar{n}}(x^{\alpha} + \delta x^{\alpha}) \hspace{3cm}$$ $$ \hspace{5cm} = \pmatrix{ \delta^{\mu}_{\sigma} && - k \delta x_{\rho} \delta^{\rho}_{\sigma}\delta^{\mu}_{\tau} \cr 0 && \delta^{\mu}_{\sigma} \delta^{\nu}_{\tau} } \pmatrix{v^{\sigma}(x^{\alpha} + \delta x^{\alpha}) \cr T^{\sigma \tau}(x^{\alpha} + \delta x^{\alpha}) } \quad    $$ \vspace{0.25cm}
$$ \hspace{5cm} =  \pmatrix{v^{\mu}(x^{\alpha}) + \left( \frac{\partial{v^{\nu}}}{\partial{x^{\sigma}}} - kT^{\mu}_{\sigma} \right) \delta x^{\sigma} + {\mathrm{O}}(\delta x^2) \cr \cr T^{\mu \nu}(x^{\alpha} + \delta x^{\alpha}) } \quad   , $$
where the Lorentz transformation is the identity $\Lambda$ = 1.Thus, to first order in $\delta x,$ the vector field changes by an amount $\delta v^{\mu}$ given by
\begin{equation} \label{Dpsi3}
\delta v^{\mu} =  \left( \frac{\partial{ v^{\mu}}}{\partial{x^{\sigma}}} - k T^{\mu}_{\sigma} \right) \delta x^{\sigma} \quad ,
\end{equation}
where the indices are lowered with the metric, $T^{\mu}_{\sigma} \equiv$ $\eta_{\rho \sigma} T^{\rho \mu}.$ The quantity $k T^{\mu}_{\sigma}$ is called a `nonlinear connection' \cite{vanDrie} because the term $k T^{\mu}_{\sigma}$ has no factor of $v^{\rho}$ and a linear connection term would have such a factor, by definition. Herein the quantity $k T^{\mu}_{\sigma}$ is called a `Poincar\'{e} connection'.

Let $X^{\mu}(\tau)$ be a suitably differentiable curve in spacetime parametized by arc length $\tau$ measured along the curve. Define $ V^{\mu}(\tau)$ to be the vector field restricted to the curve, $ V^{\mu}(\tau) \equiv $ $v^{\mu}(X(\tau)). $ By (\ref{Dpsi3}), translation along a short segment of the curve changes the vector field by $\delta V^{\nu}(\tau)$, where 
\begin{equation} \label{CURVE4}
 \delta V^{\mu} =   \left( \frac{d{ V^{\mu}}}{d \tau} - k T^{\mu}_{\sigma} \frac{dX^{\sigma}}{d\tau} \right) \delta \tau \quad .
\end{equation}
The change in the vector field due to translation along a curve has two terms because the coordinates of an event change and the components of the field $\psi$ are scrambled.
 
When $V^{\mu}(\tau)$ is {\it{parallel translated}} along the curve $X^{\nu}(\tau),$ the change (\ref{CURVE4}) vanishes, $\delta V^{\mu}(\tau)$ = 0, and it follows that
\begin{equation} \label{Ptrans1}
     \frac{d{ V^{\mu}}}{d \tau} =  k T^{\mu}_{\sigma} \frac{dX^{\sigma}}{d\tau}  \quad .
\end{equation}
A {\it{geodesic curve}} is defined to be a curve $X^{\nu}(\tau) $ which has a parallel translated tangent. Thus the vector $V^{\mu}(\tau)$ in (\ref{Ptrans1}) is taken to be the tangent vector, $ V^{\mu}$ = $dX^{\mu}/d\tau,$ yielding the geodesic equation:
\begin{equation} \label{ARC1}
     \frac{d^2{ X^{\mu}}}{d \tau^{2}} = k T^{\mu}_{\sigma} \frac{dX^{\sigma}}{d\tau} \quad .
\end{equation}
Along a geodesic, the 4-acceleration is the scalar product of the connection tensor and the tangent 4-vector. (`Geodesic curve' herein means just a curve with a parallel translated tangent, not extreme arc length.)

Dividing both sides of the arc length formula (\ref{ARC2}) by the square of the arc length shows that the square of the magnitude of the vector $dx^{\mu}/d{\tau}$ is constant. Hence the tangent vector $dX^{\mu}/d\tau$ has a constant magnitude$$
     0 =  \frac{d}{d\tau} \left(\eta_{\alpha \beta} \frac{dX^{\alpha}}{d\tau}\frac{dX^{\beta}}{d\tau} \right)=  2\eta_{\alpha \beta} \frac{dX^{\alpha}}{d\tau}\frac{d^2X^{\beta}}{d\tau^2} 
$$  
and the geodesic equation (\ref{ARC1}) implies that
$$
	0 =  \eta_{\alpha \beta} \eta_{\sigma \rho}\frac{dX^{\alpha}}{d\tau} \frac{dX^{\sigma}}{d\tau} T^{\rho \beta} \quad .
$$
To make the symmetry plain rewrite this as follows
$$0 = \left(\eta_{\sigma \rho} \frac{dX^{\sigma}}{d\tau}\right) \left(\eta_{\alpha \beta} \frac{dX^{\alpha}}{d\tau} \right)\frac{1}{2}\left(  T^{\rho \beta} + T^{\beta \rho} \right) \quad .
$$
It is sufficient that $T^{\rho \beta}$ be antisymmetric, but one can invoke reasonable physical conditions that make antisymmetry necessary. 

If the connection tensor at an event $A$ is the same for all timelike geodesics through event $A$, then the tangent vector factors in the above expression are suitably arbitrary so that $T^{\rho \beta}$ is {\it{necessarily}} antisymmetric at $A$. Consider that a `test' particle can move at any sublight speed in any direction through event $A$ in spacetime. 

Thus consider the set of all timelike vector fields with geodesics in all timelike directions due to the one connection tensor field $T^{\rho \beta}.$ The 4-vector parts of the collection of $\psi^{(a)}$s, each 4-vector part labeled with a different value of $a,$ are restricted to be the tangent vectors of geodesics and the tensor part $T^{\alpha \beta}$ is the same tensor field for any value of $a.$ Since the 4-vector field of any one $\psi^{(a)}$ can have just one value at each spacetime event, it is important that no two geodesics of any one $\psi^{(a)}$ intersect. Hence it may be necessary to consider the field $\psi^{(a)}$ defined over a patch of spacetime rather than over all of spacetime.

These considerations require that
\begin{equation} \label{LF1}
      T^{\rho \beta} = - T^{\beta \rho} \quad ,
\end{equation}
and the connection tensor $T^{\mu \nu}$ must be antisymmetric. The geodesic equation (\ref{ARC1}) has the form of the Lorentz Force Law: the acceleration along the curve (the particle's path) is proportional to the scalar product of the tangent vector (the particle's velocity 4-vector) and a second rank antisymmetric tensor (the electromagnetic field).

To make the identification explicit, rewrite the connection tensor $T^{\mu \nu}$ as
\begin{equation} \label{LF3}
     T^{\mu \nu} =  -{\frac{q}{k m}} F^{\mu \nu}  \quad ,
\end{equation}
where  $q$ is the charge and $m$ is the mass of a particle and $F^{\mu \nu}$ is the electromagnetic field. The geodesic equation (\ref{ARC1}) is then the electrodynamic force law
\begin{equation} \label{LF4}
     m\frac{d{ V^{\mu}}}{d \tau} = q F^{\mu \sigma} V_{\sigma}  \quad ,
\end{equation}
where $V^{\mu}$ is the 4-velocity vector $dX^{\mu}/d\tau,$ $V^{\mu}$ = $dX^{\mu}/d\tau.$  

The collection of fields $\psi_{n}^{(a)}(x)$ is a collection of fields of the time-like tangent vectors to possible particle paths, each $\psi_{n}^{(a)}(x)$ in the form
\begin{equation} \label{psi2}
 \psi_{n}^{(a)}(x)  = \pmatrix{ \frac{d{ X^{(a)\mu}}}{d\tau} \cr \cr  -{\frac{q}{k m}} F^{\alpha \beta} } \quad ,
\end{equation}
where `$a$' distinguishes different paths in the electromagnetic field $F^{\mu \nu}.$ 

One can turn this around. Starting with the expression (\ref{psi2}) for $\psi_{n}^{(a)}(x),$ the transformations of a 4-vector field $v^{\mu}(x)$ and of the electromagnetic field $F^{\alpha \beta}$ determine two Poincar\'{e} representations, one of which is described in Appendix \ref{Poincare}. (The other representation translates vectors trivially and may be of interest in other applications.) Then parallel translating tangent vectors with the representation in Appendix \ref{Poincare} produces just the same paths as are allowed by the Lorentz Force Law.

%\vspace{3.cm}

\pagebreak

\appendix

\section{Poincar\'{e} Transformations of the Direct Sum Field } \label{Poincare}

This Appendix, much of the text, and some of the problems repeat material presented in an earlier, more extensive work \cite{Shurtleff2}. 

The most general transformations of a 4-vector field occur when the 4-vector field is accompanied by a tensor field.\cite{Shurtleff2} The direct sum of a 4-vector field $v^{\mu}(x)$ and a 2nd rank tensor field $T^{\alpha \beta}(x)$ can be written as a column vector,
\begin{equation} \label{psi1}
 \psi_{n}(x)  = \pmatrix{v^{\mu}(x) \cr T^{\alpha \beta}(x)} \quad ,
\end{equation}
where, for convenience, the vector and tensor are assumed to be dimensionless and the index $n$ takes values in the set $\{1, \ldots , 20 \}$ since the vector has 4 components and the tensor has 16 components. Transcribing the 16 double index values $\alpha \beta$ = $\{11,12,\ldots, 44\}$ to the 16 single index values $n$ = $\{5,6,\ldots,20\}$ is treated in the Problem set, Appendix \ref{A}. 

Any Poincar\'{e} transformation $(\Lambda,\delta x)$ can be written as a homogeneous Lorentz transformation $\Lambda$ followed by a translation through a displacement $\delta x.$ The field $\psi_{n}(x)$ changes as a differential representation that changes functions of $x$ to functions of $\Lambda x + \delta x$ and also as a square matrix representation that acts on the 20 components of $\psi_{n},$ \cite{Tung1, WeinbergFields}
\begin{equation} \label{transf2}
U(\Lambda,\delta x) \psi_{n}(x) {U}^{-1}(\Lambda,\delta x) = \sum_{\bar{n}} D^{-1}_{n \bar{n}}(\Lambda,\delta x)  \psi_{\bar{n}}(\Lambda x + \delta x) \quad ,
\end{equation}
where $D(\Lambda,\delta x)$ is the covariant nonunitary matrix representing the spacetime transformation $(\Lambda,\delta x).$ 

Let $J^{\mu \nu}$ be the angular momentum and boost matrix generators of rotations and boosts and let $P^{\mu}$ be the four momentum matrices. Then the transformation matrices are determined by the displacement $\delta x_{\mu}$ and the antisymmetric parameters $ \omega_{\mu \nu}$ of the Lorentz transformation $\Lambda,$
\begin{equation} \label{transf1}
 D(\Lambda,\delta x_{\mu})  = \exp{(- i \delta x_{\mu} P^{\mu})}\exp{( i \omega_{\mu \nu} J^{\mu \nu}/2)}  \quad .
\end{equation}

The generators can be put in block matrix form as follows
\begin{equation} \label{gen1}
 J^{\rho \sigma}  = \pmatrix{(J^{\rho \sigma}_{11})^{\mu}_{\nu} && 0 \cr 0 && (J^{\rho \sigma}_{22})^{\alpha \beta}_{\gamma \delta}} \quad {\mathrm{and}} \quad P^{\mu}  = \pmatrix{0 && (P^{\mu}_{12})^{\nu}_{\alpha \beta} \cr 0 && 0}\quad ,
\end{equation}
where the 11-block generators are herein taken to be 
$$ 
 (J^{\rho \sigma}_{11})^{\mu}_{\nu}  = i \left(\eta^{\sigma \mu} \delta^{\rho}_{\nu} - \eta^{\rho \mu} \delta^{\sigma}_{\nu} \right) \quad ,
$$ 
where $\eta^{\alpha \beta}$ is the flat spacetime metric, the diagonal matrix $\eta$ = diag$\{+1,+1,+1,-1\}$ and $\delta^{\rho}_{\nu}$ is the Kronecker delta function, one for $\rho$ = $\nu$ and zero otherwise. The 22-block generators are
$$ 
 (J^{\rho \sigma}_{22})^{\gamma \delta}_{\epsilon \xi}  = -i \left(\eta^{\rho \gamma} \delta^{\sigma}_{\epsilon} \delta^{\delta}_{\xi}- \eta^{\sigma \gamma} \delta^{\rho}_{\epsilon} \delta^{\delta}_{\xi} + \eta^{\rho \delta} \delta^{\sigma}_{\xi} \delta^{\gamma}_{\epsilon}- \eta^{\sigma \delta} \delta^{\rho}_{\xi} \delta^{\gamma}_{\epsilon} \right) \quad .
$$ 
The formulas for $J^{\mu \nu}_{11}$ and $J^{\mu \nu}_{22}$ are standard formulas, taken from the literature.\cite{WeinbergJ}

A convenient formula for the 12-block components in $P^{\mu}$ is given by 
\begin{equation} \label{gen2}
 (P^{\mu}_{12})^{\nu}_{\alpha \beta}  = i [ \left( C_1 + C_2 - C_3 \right)\delta^{\mu}_{\alpha} \delta^{\nu}_{\beta}  + 
 \left( C_1 + C_2 + C_3 \right)\delta^{\mu}_{\beta} \delta^{\nu}_{\alpha}  -  2 C_1 \eta^{\mu \nu} \eta_{\alpha \beta}
  + C_4  \eta^{\mu \sigma} \eta^{\nu \rho} \epsilon_{\sigma \rho \alpha \beta}]  \quad ,
\end{equation} 
where the constants $C_{i}$ have dimensions of an inverse distance and can be chosen arbitrarily. These constants appear because the Poincar\'{e} commutation rules are homogeneous in $P^{\mu},$ so the momentum matrices are determined within a scale factor. There are four such constants because the transformations of a 2nd rank tensor combine four spins. The representation used in the text has $C_{2}$ = $-C_{3}$ = $k/2$ and $C_{1}$ = $C_{4}$ = 0, where $k$ is a constant with the dimensions of an inverse length,
$$ 
 (P^{\mu}_{12})^{\nu}_{\alpha \beta}  = i k \delta^{\mu}_{\alpha} \delta^{\nu}_{\beta}   \quad . $$ 
Expressions for the components of momentum matrices are not readily available from the literature. They can be obtained by solving the Poincar\'{e} commutation rules given the expressions above for $J^{\mu \nu}$ or from the formulas in reference \cite{Shurtleff1}. 

The transformations determined by the generators $J^{\mu \nu}$ and $P^{\mu}$ represent the Poincar\'{e} group faithfully, see the problem set in Appendix \ref{A}.

\pagebreak

\section{On Site Verbal Explanations at the Poster Session} \label{Verbal}

People who were interested in the poster knew (i) that vectors changed upon rotation and (ii) that the rotation can be represented with matrices. I believe they were thinking of something like the following, for 2-dimensions,
$$Rv = \pmatrix{\cos{\theta} && -\sin{\theta} \cr \sin{\theta} && \cos{\theta} } \pmatrix{v^{1} \cr v^{2}} = \pmatrix{v^{1} \cos{\theta} - v^{2} \sin{\theta} \cr  v^{1} \sin{\theta} + v^{2} \cos{\theta}}  \quad .$$ 
They also knew (iii) that vectors are invariant under translations, yet translations cause coordinates to change.

Upon this foundation, I pointed out to them that coordinates rotate much like vectors and the transformation of coordinates can be put in matrix form. I scrawled the following on a blank portion of poster:
$$ \pmatrix{R && \delta x \cr 0 && 1 } \pmatrix{x \cr 1} = \pmatrix{R x + \delta x \cr  1}  \quad .$$ 
I told them that the `1' under the $x$ was a scalar since it remained unchanged by the rotation and translation. 

The punchline: A vector must be transformed along with some other quantity in order to change under a translation.

Then I pointed to the 16 terms $T$ in the expression for $\psi,$ see Appendix \ref{VTM}. This is the most general quantity that can be combined with a vector to cause the vector to change with a translation. The people looked unpleased, but accepting. Quickly, so as not to lose them, they were told that the 2nd rank tensor turned out to be just the electromagnetic field, within a constant factor. 

The expression in Appendix \ref{Bf} for the case of a constant magnetic field was the next stop, showing that the translation matrix changes the $x$ and the $y$ components of velocity. Backing up to Figure 1, they were shown that the velocity can depend on position. Finally, the position dependence can be adjusted to just cancel the effects of the matrix translation. 

The punchline: The position dependence and the effects of the translation matrix can cancel so that the velocity at the second event is as `parallel' as possible to the velocity at the first event. 

Thus the Lorentz Force Law produces paths with constant velocity. I think I also sometimes said `velocity is invariant under small translations along the path.' 

One asked if there was some new phenomena here. No, its a way to interpret the well-established Lorentz force law. 

They were kind and patient and asked questions.

\pagebreak
\section{Vector-Tensor Field and Translation Matrix} \label{VTM}
%\pagebreak
$$\psi_{n}(x) = {\scriptsize{\pmatrix{ v^1(x) \cr v^2(x) \cr v^3(x) \cr v^4(x) \cr 
    T^{12}(x) \cr T^{13}(x) \cr T^{14}(x) \cr T^{23}(x) \cr T^{24}(x) \cr T^{34}(x) \cr T^{11}(x) \cr T^{22}(x) \cr T^{33}(x) \cr T^{44}(x) \cr T^{21}(x)  \cr T^{31}(x) \cr 
   T^{41}(x) \cr T^{32}(x) \cr T^{42}(x) \cr T^{43}(x)} }} \hspace{4cm} \psi_{n}(x) = {\scriptsize{\pmatrix{ v^1 \cr v^2 \cr v^3 \cr v^4 \cr 
    -\frac{q F^{12}}{k m} \cr -\frac{q F^{13}}{k m} \cr -\frac{q F^{14}}{k m} \cr -\frac{q F^{23}}{k m} \cr -\frac{q F^{24}}{k m} \cr -\frac{q F^{34}}{k m} \cr 0 \cr 0 \cr 0 \cr 0 \cr -\frac{q F^{21}}{k m}     \cr -\frac{q F^{31}}{k m} \cr 
    -\frac{q F^{41}}{k m} \cr -\frac{q F^{32}}{k m} \cr -\frac{q F^{42}}{k m} \cr -\frac{q F^{43}}{k m}} }} = {\scriptsize{\pmatrix{ v^1 \cr v^2 \cr v^3 \cr v^4 \cr 
    -\frac{q B^3}{k m} \cr \frac{q B^2}{k m} \cr \frac{q E^1}{k m} \cr -\frac{q B^1}{k m} \cr \frac{q E^2}{k m} \cr \frac{q E^3}{k m} \cr 0 \cr 0 \cr 0 \cr 0 \cr    \frac{q B^3}{k m}     \cr -\frac{q B^2}{k m} \cr 
    -\frac{q E^1}{k m} \cr \frac{q B^1}{k m} \cr -\frac{q E^2}{k m} \cr -\frac{q E^3}{k m}} \hspace{2cm} }}$$   

{\it{The general field $\psi_{n}(x)$}} (above, left) combines a 4-vector field $v^{\mu}(x)$ and a 2nd rank tensor field $T^{\alpha \beta}(x).$ It is shown that for $v^{\mu}(x)$ to be a field of tangents to geodesics, $T^{\alpha \beta}(x)$ must be antisymmetric. When $T^{\alpha \beta}(x)$ is proportional to the electromagnetic field $F^{\alpha \beta}(x),$ see above right, the geodesics are possible paths for a particle of charge $q$ and mass $m.$

$${\tiny{\pmatrix{ 1 & 0 & 0 & 0 & 0 & 0 & 0 & 0 & 0 & 0 & -   k\,
      {  {\delta x}} & 0 & 0 & 0 & - k\,{{\delta y}}& -k\,{{\delta z}} & k\,
    {  {\delta t}} & 0 & 0 & 0 \cr
 0 & 1 & 0 & 0 & -   k\, {  {\delta x}} & 0 & 0 & 0 & 0 & 0 & 0 & -  
      k\,{  {\delta y}}     & 0 & 0 & 0 & 0 & 0 & -k\, {{\delta z}} & k\, {  {\delta t}} & 0 \cr 
0 & 0 & 1 & 0 & 0 & -   k\, {  {\delta x}} & 0 & -   k\, {  {\delta y}}     & 0 & 0 & 0 & 0 & -k\, {{\delta z}} & 0 & 0 & 0 & 
    0 & 0 & 0 & k\,{  {\delta t}} \cr 
0 & 0 & 0 & 1 & 0 & 0 & - k\,{  {\delta x}} & 0 & - k\, {  {\delta y}} & -k\, {{\delta z}} & 0 & 0 & 0 & k\,
    {  {\delta t}} & 0 & 0 & 0 & 0 & 0 & 0 \cr 0 & 0 & 0 & 0 & 1 &
    0 & 0 & 0 & 0 & 0 & 0 & 0 & 0 & 0 & 0 & 0 & 0 & 0 & 0 & 0 \cr 0 & 0 & 
    0 & 0 & 0 & 1 & 0 & 0 & 0 & 0 & 0 & 0 & 0 & 0 & 0 & 0 & 0 & 0 & 0 & 
    0 \cr 0 & 0 & 0 & 0 & 0 & 0 & 1 & 0 & 0 & 0 & 0 & 0 & 0 & 0 & 0 & 0 & 
    0 & 0 & 0 & 0 \cr 0 & 0 & 0 & 0 & 0 & 0 & 0 & 1 & 0 & 0 & 0 & 0 & 0 & 
    0 & 0 & 0 & 0 & 0 & 0 & 0 \cr 0 & 0 & 0 & 0 & 0 & 0 & 0 & 0 & 1 & 0 & 
    0 & 0 & 0 & 0 & 0 & 0 & 0 & 0 & 0 & 0 \cr 0 & 0 & 0 & 0 & 0 & 0 & 0 & 
    0 & 0 & 1 & 0 & 0 & 0 & 0 & 0 & 0 & 0 & 0 & 0 & 0 \cr 0 & 0 & 0 & 0 & 
    0 & 0 & 0 & 0 & 0 & 0 & 1 & 0 & 0 & 0 & 0 & 0 & 0 & 0 & 0 & 0 \cr 0 & 
    0 & 0 & 0 & 0 & 0 & 0 & 0 & 0 & 0 & 0 & 1 & 0 & 0 & 0 & 0 & 0 & 0 & 0 & 
    0 \cr 0 & 0 & 0 & 0 & 0 & 0 & 0 & 0 & 0 & 0 & 0 & 0 & 1 & 0 & 0 & 0 & 
    0 & 0 & 0 & 0 \cr 0 & 0 & 0 & 0 & 0 & 0 & 0 & 0 & 0 & 0 & 0 & 0 & 0 & 
    1 & 0 & 0 & 0 & 0 & 0 & 0 \cr 0 & 0 & 0 & 0 & 0 & 0 & 0 & 0 & 0 & 0 & 
    0 & 0 & 0 & 0 & 1 & 0 & 0 & 0 & 0 & 0 \cr 0 & 0 & 0 & 0 & 0 & 0 & 0 & 
    0 & 0 & 0 & 0 & 0 & 0 & 0 & 0 & 1 & 0 & 0 & 0 & 0 \cr 0 & 0 & 0 & 0 & 
    0 & 0 & 0 & 0 & 0 & 0 & 0 & 0 & 0 & 0 & 0 & 0 & 1 & 0 & 0 & 0 \cr 0 & 
    0 & 0 & 0 & 0 & 0 & 0 & 0 & 0 & 0 & 0 & 0 & 0 & 0 & 0 & 0 & 0 & 1 & 0 & 
    0 \cr 0 & 0 & 0 & 0 & 0 & 0 & 0 & 0 & 0 & 0 & 0 & 0 & 0 & 0 & 0 & 0 & 
    0 & 0 & 1 & 0 \cr 0 & 0 & 0 & 0 & 0 & 0 & 0 & 0 & 0 & 0 & 0 & 0 & 0 & 
    0 & 0 & 0 & 0 & 0 & 0 & 1 \cr  } }}$$

{\it{The translation matrix}} above acts on the components of $\psi_{n}(x)$ much like the rotation matrix $\pmatrix{\cos{\theta} && \sin{\theta} \cr - \sin{\theta} && \cos{\theta} }$ acts on a 2-dimensional vector in the Euclidean plane or a spin 1/2 rotation matrix acts on a spin 1/2 wave function. The translation matrix is far from arbitrary; it must work alongside rotation and boost matrices.

\pagebreak
\section{Constant Magnetic Field} \label{Bf}

\vspace{2cm}

\begin{figure}[h] \label{fig1}	% 1 CM.ps from pathfig.nb
\vspace{0in}
\hspace{0in}\includegraphics[0,0][288,288]{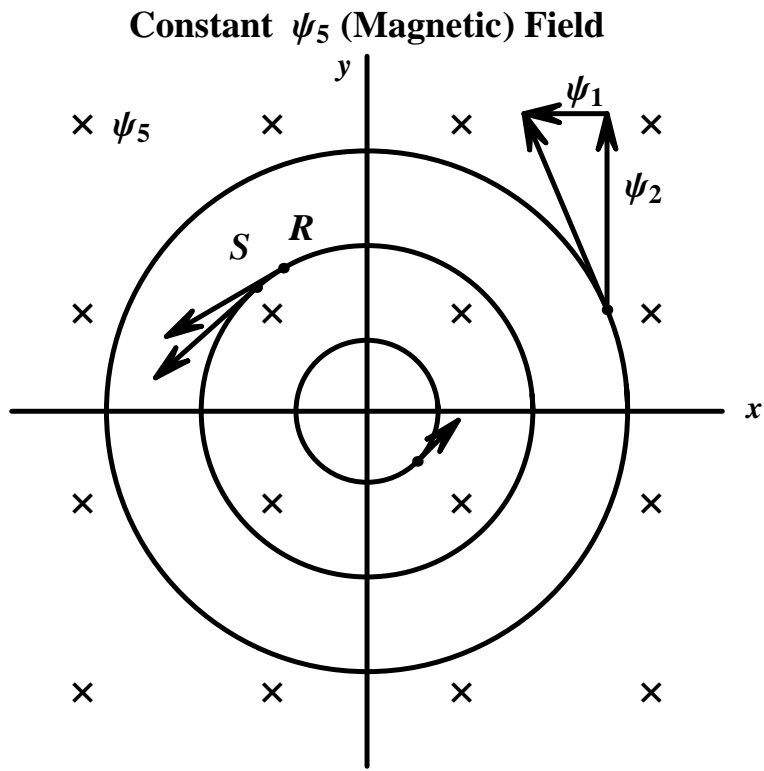}
\caption{ {\it{Translation changes the field because the coordinates change.}} The $x$-velocity at event $S,$ $\psi_{1}(S),$ differs from the $x$-velocity at $R,$ $\psi_{1}(R).$ Translating the field $\psi$ from $R$ to $S$ remaps the velocity field so that the new value at $R$ is the old value at $S$. The change in $\psi_{1}(R)$ to first order is just $ \partial_{x}\psi_{1} \delta x +\partial_{y}\psi_{1} \delta y,$ where $\delta x$ is the $x$-component of $RS.$  }
\end{figure}

\pagebreak

\vspace*{1cm}

{\centerline{{\bf{Translation Matrix}} $\cdot$ {\bf{Field}} \hspace{0.25cm} = \hspace{0.25cm} {\bf{Translated Field}}}}

$${\tiny{\pmatrix{ 1 & 0 & 0 & 0 & 0 & 0 & 0 & 0 & 0 & 0 & -   k\,
      {  {\delta x}}     & 0 & 0 & 0 & -   k\,
      {  {\delta y}}     & 0 & k\,
    {  {\delta t}} & 0 & 0 & 0 \cr 0 & 1 & 0 & 0 & -   k\,
      {  {\delta x}}     & 0 & 0 & 0 & 0 & 0 & 0 & -  
      k\,{  {\delta y}}     & 0 & 0 & 0 & 0 & 0 & 0 & k\,
    {  {\delta t}} & 0 \cr 0 & 0 & 1 & 0 & 0 & -   k\,
      {  {\delta x}}     & 0 & -   k\,
      {  {\delta y}}     & 0 & 0 & 0 & 0 & 0 & 0 & 0 & 0 & 
    0 & 0 & 0 & k\,{  {\delta t}} \cr 0 & 0 & 0 & 1 & 0 & 0 & 
     -   k\,{  {\delta x}}     & 0 & -   k\,
      {  {\delta y}}     & 0 & 0 & 0 & 0 & k\,
    {  {\delta t}} & 0 & 0 & 0 & 0 & 0 & 0 \cr 0 & 0 & 0 & 0 & 1 &
    0 & 0 & 0 & 0 & 0 & 0 & 0 & 0 & 0 & 0 & 0 & 0 & 0 & 0 & 0 \cr 0 & 0 & 
    0 & 0 & 0 & 1 & 0 & 0 & 0 & 0 & 0 & 0 & 0 & 0 & 0 & 0 & 0 & 0 & 0 & 
    0 \cr 0 & 0 & 0 & 0 & 0 & 0 & 1 & 0 & 0 & 0 & 0 & 0 & 0 & 0 & 0 & 0 & 
    0 & 0 & 0 & 0 \cr 0 & 0 & 0 & 0 & 0 & 0 & 0 & 1 & 0 & 0 & 0 & 0 & 0 & 
    0 & 0 & 0 & 0 & 0 & 0 & 0 \cr 0 & 0 & 0 & 0 & 0 & 0 & 0 & 0 & 1 & 0 & 
    0 & 0 & 0 & 0 & 0 & 0 & 0 & 0 & 0 & 0 \cr 0 & 0 & 0 & 0 & 0 & 0 & 0 & 
    0 & 0 & 1 & 0 & 0 & 0 & 0 & 0 & 0 & 0 & 0 & 0 & 0 \cr 0 & 0 & 0 & 0 & 
    0 & 0 & 0 & 0 & 0 & 0 & 1 & 0 & 0 & 0 & 0 & 0 & 0 & 0 & 0 & 0 \cr 0 & 
    0 & 0 & 0 & 0 & 0 & 0 & 0 & 0 & 0 & 0 & 1 & 0 & 0 & 0 & 0 & 0 & 0 & 0 & 
    0 \cr 0 & 0 & 0 & 0 & 0 & 0 & 0 & 0 & 0 & 0 & 0 & 0 & 1 & 0 & 0 & 0 & 
    0 & 0 & 0 & 0 \cr 0 & 0 & 0 & 0 & 0 & 0 & 0 & 0 & 0 & 0 & 0 & 0 & 0 & 
    1 & 0 & 0 & 0 & 0 & 0 & 0 \cr 0 & 0 & 0 & 0 & 0 & 0 & 0 & 0 & 0 & 0 & 
    0 & 0 & 0 & 0 & 1 & 0 & 0 & 0 & 0 & 0 \cr 0 & 0 & 0 & 0 & 0 & 0 & 0 & 
    0 & 0 & 0 & 0 & 0 & 0 & 0 & 0 & 1 & 0 & 0 & 0 & 0 \cr 0 & 0 & 0 & 0 & 
    0 & 0 & 0 & 0 & 0 & 0 & 0 & 0 & 0 & 0 & 0 & 0 & 1 & 0 & 0 & 0 \cr 0 & 
    0 & 0 & 0 & 0 & 0 & 0 & 0 & 0 & 0 & 0 & 0 & 0 & 0 & 0 & 0 & 0 & 1 & 0 & 
    0 \cr 0 & 0 & 0 & 0 & 0 & 0 & 0 & 0 & 0 & 0 & 0 & 0 & 0 & 0 & 0 & 0 & 
    0 & 0 & 1 & 0 \cr 0 & 0 & 0 & 0 & 0 & 0 & 0 & 0 & 0 & 0 & 0 & 0 & 0 & 
    0 & 0 & 0 & 0 & 0 & 0 & 1 \cr  } \pmatrix{ v^1 \cr v^2 \cr 0 \cr v^4 \cr 
    \frac{q B}{k m} \cr 0 \cr 0 \cr 0 \cr 0 \cr 0 \cr 0 \cr 0 \cr 0 \cr 0 \cr -   \frac{q B}{k m}     \cr 0 \cr 
    0 \cr 0 \cr 0 \cr 0} }}$$ $$\hspace{12cm} {\tiny{ =  
\pmatrix{ v^1 + \frac{q B \delta y }{m}\cr v^2 - \frac{q B \delta x }{m} \cr 0 \cr v^4 \cr 
    \frac{q B}{k m} \cr 0 \cr 0 \cr 0 \cr 0 \cr 0 \cr 0 \cr 0 \cr 0 \cr 0 \cr -   \frac{q B}{k m}     \cr 0 \cr 
    0 \cr 0 \cr 0 \cr 0}  }}\quad$$ 

\vspace{0.5cm}

{\it{Translation changes the field another way.}} Because the translation matrix acts linearly on the field components, the $x$-velocity $\psi_{1}(R)$ at $R$ changes by a term $-k \psi_{15} \delta y$ = $-qB^{3} \delta y / m$ = $+qB \delta y / m.$ The field $\psi$ includes the electromagnetic field as well as the 4-velocity. For the ordering assumed here this means that the components $\psi_{5}$ and $\psi_{15}$ are nonzero. 

The change to the $x$-velocity $\psi_{1}(R)$ occurs because the translation matrix has a nonzero component at the fifteenth column of the first row. And the fifteenth column of the first row is nonzero because of spacetime symmetry, i.e. the Poincar\'{e} commutation rules, and the specific matrices assumed for rotations and boosts.

\pagebreak

\vspace*{2cm}

\begin{figure}[h] \label{fig2}	% 1 CM.ps from pathfig.nb
\hspace{0in}\includegraphics[0,0][288,288]{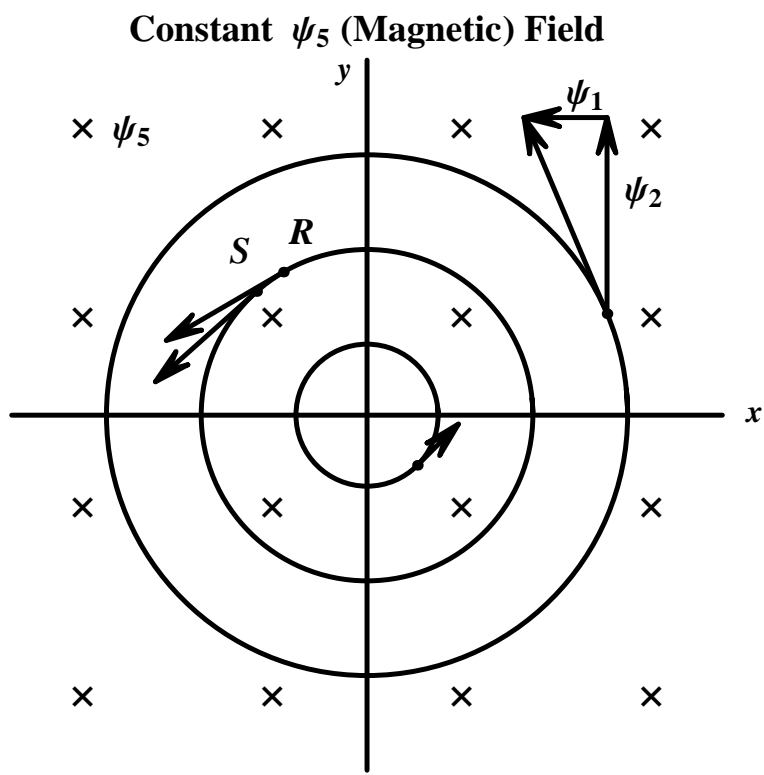}
%\vspace{7.cm}
%\hskip1.75in\special{eps:c:/PCTeXv4/00-07tex/SENDxxx/fig1.eps x=7cm y=7.cm}
\caption{ {\it{Translation invariance.}} Thus there are two ways that $\psi$ changes with a translation. Curves for which the two effects cancel are called `geodesics'. The sketch shows three geodesics for a constant $\psi_{5}$ = $-\psi_{15}$ field, i.e. a constant magnetic field. The translation from $R$ to $S$ of $\psi$ does not change the value of $\psi$ at $R$.
The observed paths of charged particles in electromagnetic fields are just the paths that have translation invariant 4-velocities. }
\end{figure}

\pagebreak

\section{Constant Electric Field} \label{Ef}

\vspace*{2cm}

\begin{figure}[h] \label{fig3}	% 1 CM.ps from pathfig.nb
\vspace{0in}
\hspace{0in}\includegraphics[0,0][288,288]{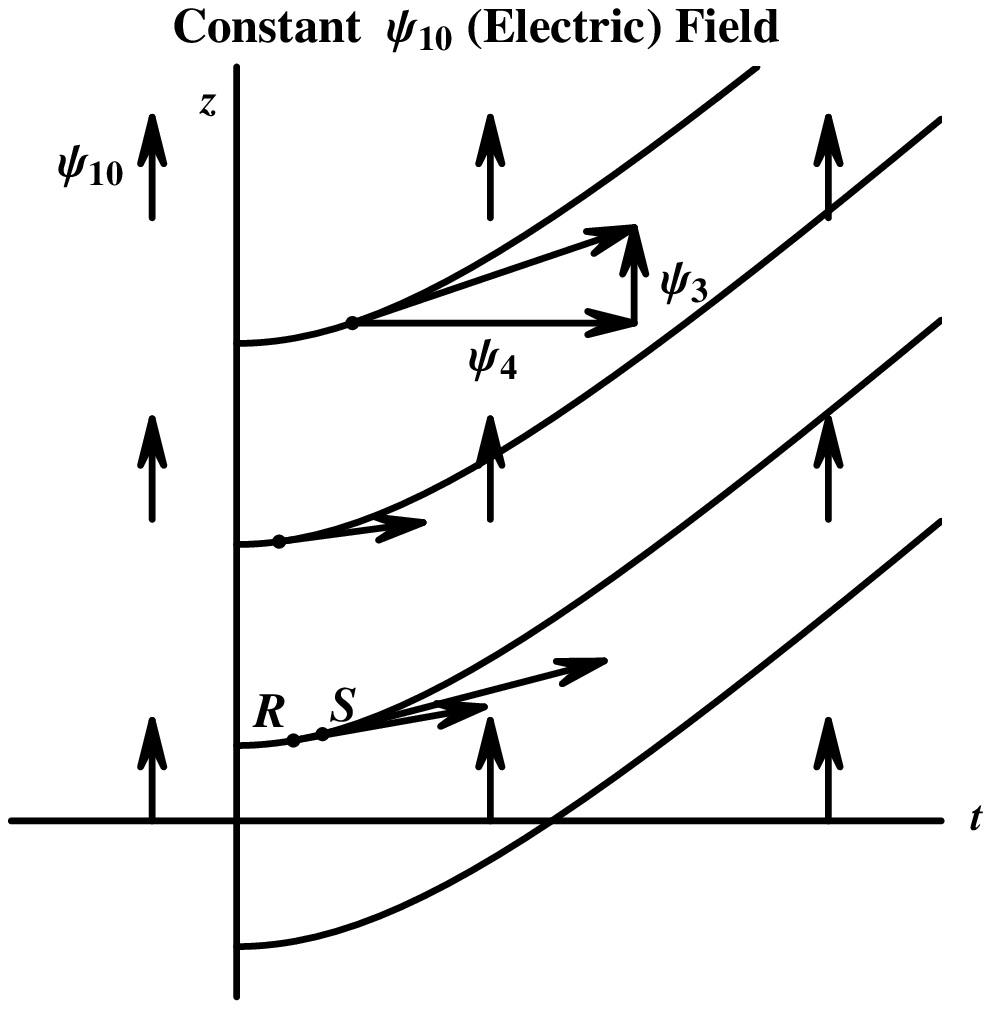}
\caption{ {\it{Translation changes the field because the coordinates change.}} The curves show the paths of particles starting at rest on the $z$-axis at $t$ = 0 and accelerating in the $z$-direction. The $z$-velocity at event $S,$ $\psi_{3}(S),$ is larger than the $z$-velocity at $R,$ $\psi_{3}(R).$ The time dilation factor $dt/d\tau$ is the fourth component of $\psi,$ $\psi_{4}.$ This component also increases from $R$ to $S,$ so a translation along $RS$ maps larger values onto $\psi_{3}(R)$ and $\psi_{4}(R)$ at $R.$ }
\end{figure}

\pagebreak

\vspace*{1cm}

{\centerline{{\bf{Translation Matrix}} $\cdot$ {\bf{Field}} \hspace{0.25cm} = \hspace{0.25cm} {\bf{Translated Field}}}}

$${\scriptsize{\pmatrix{ 1 & 0 & 0 & 0 & 0 & 0 & 0 & 0 & 0 & 0 & 0 & 0 & 0 & 0 & 0 & -   k \delta z
          & k
    \delta t & 0 & 0 & 0 \cr 0 & 1 & 0 & 0 & 0 & 0 & 0 & 0 & 0 & 0 & 0 & 0 & 0 & 0 & 0 & 0 & 
    0 & -   k{\delta z}     & k
    {\delta t} & 0 \cr 0 & 0 & 1 & 0 & 0 & 0 & 0 & 0 & 0 & 0 & 0 & 0 & -   k
      {\delta z}     & 0 & 0 & 0 & 0 & 0 & 0 & k
    {\delta t} \cr 0 & 0 & 0 & 1 & 0 & 0 & 0 & 0 & 0 & -   k{\delta z}
          & 0 & 0 & 0 & k
    {\delta t} & 0 & 0 & 0 & 0 & 0 & 0 \cr 0 & 0 & 0 & 0 & 1 & 0 & 0 & 0 & 0 & 0 & 0 & 0 & 0 & 
    0 & 0 & 0 & 0 & 0 & 0 & 0 \cr 0 & 0 & 0 & 0 & 0 & 1 & 0 & 0 & 0 & 0 & 0 & 0 & 0 & 0 & 0 & 0 & 0 & 0 & 0 &
    0 \cr 0 & 0 & 0 & 0 & 0 & 0 & 1 & 0 & 0 & 0 & 0 & 0 & 0 & 0 & 0 & 0 & 0 & 0 & 0 & 0 \cr 0 & 0 & 0 & 0 & 
    0 & 0 & 0 & 1 & 0 & 0 & 0 & 0 & 0 & 0 & 0 & 0 & 0 & 0 & 0 & 0 \cr 0 & 0 & 0 & 0 & 0 & 0 & 0 & 0 & 1 & 0 &
    0 & 0 & 0 & 0 & 0 & 0 & 0 & 0 & 0 & 0 \cr 0 & 0 & 0 & 0 & 0 & 0 & 0 & 0 & 0 & 1 & 0 & 0 & 0 & 0 & 0 & 0 &
    0 & 0 & 0 & 0 \cr 0 & 0 & 0 & 0 & 0 & 0 & 0 & 0 & 0 & 0 & 1 & 0 & 0 & 0 & 0 & 0 & 0 & 0 & 0 & 0 \cr 0 & 
    0 & 0 & 0 & 0 & 0 & 0 & 0 & 0 & 0 & 0 & 1 & 0 & 0 & 0 & 0 & 0 & 0 & 0 & 0 \cr 0 & 0 & 0 & 0 & 0 & 0 & 0 &
    0 & 0 & 0 & 0 & 0 & 1 & 0 & 0 & 0 & 0 & 0 & 0 & 0 \cr 0 & 0 & 0 & 0 & 0 & 0 & 0 & 0 & 0 & 0 & 0 & 0 & 0 &
    1 & 0 & 0 & 0 & 0 & 0 & 0 \cr 0 & 0 & 0 & 0 & 0 & 0 & 0 & 0 & 0 & 0 & 0 & 0 & 0 & 0 & 1 & 0 & 0 & 0 & 0 &
    0 \cr 0 & 0 & 0 & 0 & 0 & 0 & 0 & 0 & 0 & 0 & 0 & 0 & 0 & 0 & 0 & 1 & 0 & 0 & 0 & 0 \cr 0 & 0 & 0 & 0 & 
    0 & 0 & 0 & 0 & 0 & 0 & 0 & 0 & 0 & 0 & 0 & 0 & 1 & 0 & 0 & 0 \cr 0 & 0 & 0 & 0 & 0 & 0 & 0 & 0 & 0 & 0 &
    0 & 0 & 0 & 0 & 0 & 0 & 0 & 1 & 0 & 0 \cr 0 & 0 & 0 & 0 & 0 & 0 & 0 & 0 & 0 & 0 & 0 & 0 & 0 & 0 & 0 & 0 &
    0 & 0 & 1 & 0 \cr 0 & 0 & 0 & 0 & 0 & 0 & 0 & 0 & 0 & 0 & 0 & 0 & 0 & 0 & 0 & 0 & 0 & 0 & 0 & 1 \cr  } 
    \pmatrix{ 0 \cr 0 \cr v^3 \cr v^4 \cr 0 \cr 0 \cr 0 \cr 0 \cr 0 \cr \frac{q E}{km} \cr 0 \cr 0 \cr 0 \cr 0 \cr 0 \cr 0 \cr 0 \cr 0 \cr 0 \cr 
    -   \frac{q E}{km}    } }} $$ $$ \hspace{12cm} {\scriptsize{ = \pmatrix{ 0 \cr 0 \cr v^3 - \frac{q E \delta t }{m} \cr v^4 - \frac{q E \delta z }{m} \cr 0 \cr 0 \cr 0 \cr 0 \cr 0 \cr \frac{q E}{km} \cr 0 \cr 0 \cr 0 \cr 0 \cr 0 \cr 0 \cr 0 \cr 0 \cr 0 \cr 
    -\frac{q E}{km} } }}$$

\vspace{0.5cm}

{\it{Translation changes the field another way.}} In this case $\psi_{10}$ and $\psi_{20}$ are equal and opposite, corresponding to the case of a constant electric field in the $z$-direction. The translation matrix for a displacement along $RS$ applied to $\psi$ decreases the values of $\psi_{3}$ and $\psi_{4}$ everywhere.

\pagebreak

\vspace*{2cm}

\begin{figure}[h] \label{fig4}	% 1 CM.ps from pathfig.nb
\vspace{0in}
\hspace{0in}\includegraphics[0,0][288,288]{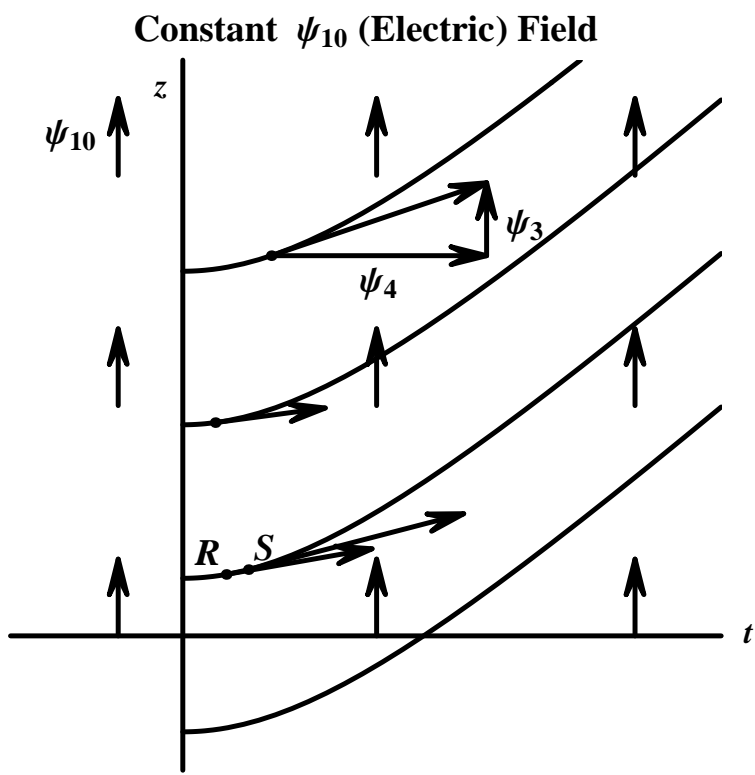}
\caption{{\it{Translation invariance.}} The increases to $\psi_{3}(R)$ and $\psi_{4}(R)$ by translation can adjusted to be just canceled by the decreases brought on by matrix translation. Each of the curves drawn above are obtained by requiring such cancelations to occur at every event along each curve. If a particle is moving along one of these curves, and its 4-velocity and the electric field make up the field $\psi,$ then translation along the curve does not change the 4-velocity. In that sense these are constant 4-velocity paths in spacetime. The Lorentz force law produces a path for a charged particle that has a 4-velocity which is left invariant by a small translation along the path. }
\end{figure}

\pagebreak

\section{Origin of the Connection: Spacetime Symmetry} \label{Origin}

Angular Momentum, Boost Matrices:
$$
 J^{\rho \sigma}  = \pmatrix{(J^{\rho \sigma}_{11})^{\mu}_{\nu} && 0 \cr 0 && (J^{\rho \sigma}_{22})^{\gamma \delta}_{\epsilon \xi}} \quad $$ 
$$
 J^{\rho \sigma}  = \pmatrix{i \left(\eta^{\sigma \mu} \delta^{\rho}_{\nu} - \eta^{\rho \mu} \delta^{\sigma}_{\nu} \right) && 0 \cr 0 && -i \left(\eta^{\rho \gamma} \delta^{\sigma}_{\epsilon} \delta^{\delta}_{\xi}- \eta^{\sigma \gamma} \delta^{\rho}_{\epsilon} \delta^{\delta}_{\xi} + \eta^{\rho \delta} \delta^{\sigma}_{\xi} \delta^{\gamma}_{\epsilon}- \eta^{\sigma \delta} \delta^{\rho}_{\xi} \delta^{\gamma}_{\epsilon} \right)} \quad .$$ 
\vspace{0.5cm}

\noindent Momentum Matrices:
$$ \quad P^{\mu}  = \pmatrix{0 && (P^{\mu}_{12})^{\nu}_{\alpha \beta} \cr 0 && 0}= \pmatrix{0 && i k \delta^{\mu}_{\alpha} \delta^{\nu}_{\beta} \cr 0 && 0}\quad .$$
\vspace{1cm}

\noindent Poincar\'{e} Commutation Rules:
$$  [J^{\mu \nu},J^{\rho \sigma}] = i [ \eta^{\mu \rho} J^{\nu \sigma} -  \eta^{\mu \sigma} J^{\nu \rho} -  \eta^{\nu \rho} J^{\mu \sigma} +  \eta^{\nu \sigma} J^{\mu \rho}] \quad ;
$$
$$[P^{\mu},J^{\rho \sigma}]  = -i \left( \eta^{\mu \rho} P^{\sigma} -  \eta^{\mu \sigma} P^{\rho}\right) \quad ; \hspace{1cm} [P^{\mu},P^{\rho }] = 0 \quad .
$$
\vspace{1cm}

\noindent Translation Matrix:
 $$ D(1,\delta x) = \exp{(-i \delta x_{\sigma} P^{\sigma})} = 1 -i \delta x_{\sigma} P^{\sigma}  = \pmatrix{\delta^{\mu}_{\nu} &&  k \delta x_{\sigma} \delta^{\sigma}_{\alpha} \delta^{\mu}_{\beta} \cr 0 && \delta^{\gamma}_{\epsilon}\delta^{\delta}_{\xi}}\quad .$$
\vspace{1cm}

\noindent Origin of the Poincar\'{e} Connection: Translation of the direct sum of a vector and a tensor 
 $$ D^{-1}_{n \bar{n}}(1,\delta x)\psi_{\bar{n}} =  \pmatrix{\delta^{\mu}_{\nu} &&  -k \delta x_{\sigma} \delta^{\sigma}_{\alpha} \delta^{\mu}_{\beta} \cr 0 && \delta^{\gamma}_{\epsilon}\delta^{\delta}_{\xi}} \pmatrix{v^{\nu} \cr  T^{\alpha \beta}} =  \pmatrix{v^{\mu} - k \delta x_{\sigma}  T^{\sigma \mu}\cr T^{\alpha \beta}} \quad .$$
The connection is the factor $k T^{\sigma \mu}$ of $\delta x_{\sigma}$ in the term added to $v^{\mu}.$

\pagebreak

\section{Parallel Translation and The Lorentz Force Law} \label{Ptrans}

Translation of the direct sum of a vector field and a tensor field:
 $$ U(\Lambda,\delta x) \psi_{n}(x) {U}^{-1}(\Lambda,\delta x) = D^{-1}_{n \bar{n}}(1,\delta x)\psi_{\bar{n}}(x^{\rho} + \delta x^{\rho} ) =    \pmatrix{v^{\mu}(x^{\rho} + \delta x^{\rho} ) - k \delta x_{\sigma}  T^{\sigma \mu}(x^{\rho} + \delta x^{\rho} )\cr T^{\alpha \beta}(x^{\rho} + \delta x^{\rho} )} $$
\vspace{0.3cm}

\noindent Change in the vector field due to the translation:
$$
\delta v^{\mu} =  \left( \frac{\partial{ v^{\mu}}}{\partial{x^{\sigma}}} - k T^{\mu}_{\sigma} \right) \delta x^{\sigma} + {\mathrm{O}}(\delta x^2)\quad .
$$
\vspace{0.3cm}

\noindent Change in the vector field when the displacement is along a curve $X(\tau)$:
$$
 \delta V^{\mu} =   \left( \frac{d{ V^{\mu}}}{d \tau} - k T^{\mu}_{\sigma} \frac{dX^{\sigma}}{d\tau} \right) \delta \tau + {\mathrm{O}}(\delta \tau^2) \quad ,
$$
where $V^{\mu}(\tau)$ = $v^{\mu}(X(\tau)).$

\vspace{0.3cm}

\noindent Geodesic Equation: Parallel translated tangent vector, $\delta V^{\mu}$ = 0 for $V^{\mu}(\tau)$ = $dX^{\mu}/d\tau,$
$$
\frac{d^2{ X^{\mu}}}{d \tau^{2}} = k T^{\mu}_{\sigma} \frac{dX^{\sigma}}{d\tau} \quad .
$$
\vspace{0.3cm}

\noindent Arc Length:
$$
     (d\tau)^2 =  -\eta_{\alpha \beta} dx^{\alpha}dx^{\beta} \quad .
$$
\vspace{0.3cm}

\noindent Geodesic curve tangent vector has a constant magnitude:
$$
     0 =  \frac{d}{d\tau} \left(\eta_{\alpha \beta} \frac{dX^{\alpha}}{d\tau}\frac{dX^{\beta}}{d\tau} \right) = \left(\eta_{\sigma \rho} \frac{dX^{\sigma}}{d\tau}\right) \left(\eta_{\alpha \beta} \frac{dX^{\alpha}}{d\tau} \right)\left(  T^{\rho \beta} + T^{\beta \rho} \right) \quad .  
$$
If true for arbitrary timelike tangents, then $T^{\rho \beta}$ is antisymmetric: $T^{\rho \beta}$ = $-T^{\beta \rho}.$

\vspace{0.3cm}

\noindent Geodesic equation with new notation for $T^{\alpha \beta}$ = $ -(e/k m) F^{\alpha \beta}$ and with $V^{\mu}$ = $dX^{\mu}/d\tau$:
$$
     m\frac{d{ V^{\mu}}}{d \tau} = e F^{\mu \sigma} V_{\sigma}  \quad ,
$$
which is the force law of electrodynamics when $m$ is the mass and $e$ is the charge of a particle in an electromagnetic field $F^{\alpha \beta}.$

\pagebreak

\section{Problems} \label{A}

\vspace{0.3cm}
\noindent 0. The representation of the Poincar\'{e} group for a direct sum field with spin $(A,B)\oplus(C,D)$ has nontrivial translation matrices when $\mid A-C \mid$ = $\mid B-D \mid$ = 1/2.\cite{Shurtleff1} The spin for a 4-vector is $(1/2,1/2).$ (a) What spins $(C,D)$ can be combined with a 4-vector in a direct sum field which has nontrivial translation matrix representation? (b) What are the values of $C$ and $D$ that correspond to the case of transforming coordinates which is displayed in Appendix \ref{Verbal}?

\vspace{0.3cm}
\noindent 1. Rewrite the geodesic equation with a new arc length $s$. That is, simplify the equation resulting from substituting the function $\tau(s)$ for $\tau$ in (\ref{ARC1}). In particular, try $\tau(s)$ =  $c_0 + c_1 s,$ where $c_{i}$ are constants.

\vspace{0.3cm}
\noindent 2. (a) Using the expressions in Appendix \ref{Poincare}, construct the 20x20 matrix generator $J^{12}$ of rotations in the 12-plane (the $xy$-plane). (b) Also construct the matrix generator of boosts in the 3 direction, $J^{34}.$  (c) Find the matrix $R^{z}(\theta)$ that represents a rotation of the 12-plane by an angle $\theta$. (d) Find the matrix that represents a boost of the coordinate system by a boost velocity of $dx^{3}/dx^{4}$ = $\tanh{(\phi)}.$ 

\noindent [Hint: To make matrices convert the 16 values of the double indices ${\gamma \delta}$ and ${\epsilon \xi}$ to single indices that range from 5 to 20 because 1 to 4 is reserved for the vector part. (One choice of ordering is $\{12,13,14,23,24,34,$ $11,22,33,44,21,31,41,32,42,43 \},$ so that a single index of $i$ = $4+5$ = 9 indicates the double index ${\gamma \delta}$ = 24.)]

\vspace{0.3cm}
\noindent 3. A charged particle in a constant magnetic field. (a) Solve the geodesic equation (\ref{ARC1}) assuming the field $T^{\alpha \beta}(x)$ has just two constant nonzero components, $T^{12}$ = $-T^{21} \neq$ 0. Describe in words a typical geodesic with $V^{3}$ = 0. (b) Write the components of the field $\psi$ for a set of non-intersecting geodesics with $V^{3}$ = 0. 

\vspace{0.3cm}
\noindent 4. A charged particle in a constant electric field. (a) Solve the geodesic equation (\ref{ARC1}) assuming the field $T^{\alpha \beta}(x)$ has just two constant nonzero components, $T^{34}$ = $-T^{43} \neq$ 0. Describe in words a typical geodesic with $V^{1}$ = $V^{2}$ = 0. (b) Write the components of the field $\psi$ for a set of non-intersecting geodesics with $V^{1}$ = $V^{2}$ = 0. 

\vspace{0.3cm}
\noindent 5. (a) Show that the 11-block matrices $J^{\alpha \beta}_{11}$ in (\ref{gen1}) satisfy the Poincar\'{e} algebra commutation rule
$$  [J^{\mu \nu}_{11},J^{\rho \sigma}_{11}] = i [ \eta^{\mu \rho} J^{\nu \sigma}_{11} -  \eta^{\mu \sigma} J^{\nu \rho}_{11} -  \eta^{\nu \rho} J^{\mu \sigma}_{11} +  \eta^{\nu \sigma} J^{\mu \rho}_{11}] \quad ,
$$
where $[J^{\mu \nu}_{11},J^{\rho \sigma}_{11}]^{\alpha}_{\gamma}$ = $(J^{\mu \nu}_{11})^{\alpha}_{\beta}  (J^{\rho \sigma}_{11})^{\beta}_{\gamma} - (J^{\rho \sigma}_{11})^{\alpha}_{\beta}  (J^{\mu \nu}_{11})^{\beta}_{\gamma}$  .

(b) Show that the momentum-angular momentum commutation rule is satisfied by the expressions in Appendix \ref{Poincare}. The [P,J] rule is trivial except for the 12-block,
$$[P^{\mu},J^{\rho \sigma}]_{12} = (P^{\mu}_{12})^{\nu}_{\alpha \beta}(J^{\rho \sigma}_{22})^{\alpha \beta}_{\gamma \delta} - (J^{\rho \sigma}_{11})^{\nu}_{\sigma} (P^{\mu}_{12})^{\sigma}_{\gamma \delta} = -i \left( \eta^{\mu \rho} P^{\sigma}_{12} -  \eta^{\mu \sigma} P^{\rho}_{12}\right) \quad .
$$

\vspace{0.3cm}
\noindent 6. In general, the momentum matrices in (\ref{gen1}) and (\ref{gen2}) produce a Poincar\'{e} connection that depends on the constants $C_{i}.$ The connection $kT^{\nu}_{\sigma}$ in the text is a special case of the more general expression $k \Gamma^{\nu}_{\sigma}$ =  $-i \eta_{\mu \sigma} (P^{\mu}_{12})^{\nu}_{\alpha \beta} T^{\alpha \beta}$ (a) Show that $kT^{\nu}_{\sigma}$ = $k \Gamma^{\nu}_{\sigma}$ for $C_{2}$ = $-C_{3}$ = $k/2$ and $C_{1}$ = $C_{4}$ = 0 . (b) For what values of $C_{i}$ is $\Gamma^{\mu \nu}$ symmetric in $\mu \nu$? (c) Antisymmetric? (d) The trace of a tensor, $\bar{T} \equiv$ $\bar{T}^{11} + \bar{T}^{22} + \bar{T}^{33} - \bar{T}^{44}$ is an invariant. Find constants $C_{i}$ such that $\Gamma^{\mu \nu}$ is $\eta^{\mu \nu}$ times the trace $T$ of the tensor part of $\psi.$

\vspace{0.3cm}
\noindent 7. Repeat the arguments presented in the text, but for a position-dependent metric $g_{\alpha \beta}(x),$ making reasonable assumptions whenever the need arises. [See reference \cite{Shurtleff2}.]

%\pagebreak

\end{document}